\documentclass[ reprint,nofootinbib,amsmath,amssymb,aps]{revtex4-2}
\usepackage{appendix}
\usepackage{graphicx}
\usepackage{dcolumn}
\newcommand{\rp}{r_+}
\renewcommand{\rm}{r_-}
\usepackage[colorlinks,linkcolor=magenta,anchorcolor=cyan,citecolor=blue]{hyperref}
\usepackage{physics}
\usepackage{bm}
\usepackage{slashed}
\usepackage{mathrsfs}
\usepackage[multiple]{footmisc}

\begin{document}
\title{ Notes on solution phase space and BTZ black hole}
\author{Wei Guo$^{1,2}$}
\email{guow@mail.bnu.edu.cn}
\affiliation{$^1$School of Physics and Astronomy, Beijing Normal University, Beijing 100875, China\label{addr1}\\
$^2$ Key Laboratory of Multiscale Spin Physics, Ministry of Education, Beijing Normal University, Beijing 100875, China}

\date{\today}

\begin{abstract}
In this paper, we use the solution phase space approach based on the covariant phase space formalism to compute the conserved charges of the BTZ black hole, namely mass, angular momentum, and entropy. Furthermore, we discuss the first law of the BTZ black hole and the Smarr relation. For completeness, outer horizon and inner horizon cases have been all included. Additionally, the results of the three-dimensional Kerr-dS spacetime have also been obtained. Our results are consistent with previous investigations. Considering the simplicity of the circumstances, we have presented the most detailed possible information, with the aim of facilitating research in related fields.
\end{abstract}
\maketitle
\section{Introduction}
\label{sec:intro}

Symmetry is fundamental in modern physics, appearing in various forms, notably as the symmetry of a theory and the symmetry of a specific solution. In gravitational physics, symmetries of solutions, known as isometries, are represented by Killing vectors. A symmetry in a theory, determined by its action and initial/boundary conditions, remains invariant under transformations that map one solution to another. These transformations can be discrete or continuous, with the latter often described by Lie groups. Continuous symmetries are classified as either global or local, with local symmetries allowing transformations based on arbitrary spacetime functions, leading to gauge theories. In such theories, different solutions can be related through local symmetries, and physical theories assume these solutions are equivalent, forming gauge classes. Gauge symmetries, therefore, represent redundant degrees of freedom. Emmy Noether's theorems link symmetries with conservation laws\cite{Noether:1918zz}, with the first theorem associating conserved charges to symmetries and the second constraining field equations in gauge theories. 

Following the foundational work of Emmy Noether, the relationship between symmetries and conserved charges has become central to theoretical physics. However, in General Relativity (GR), defining these charges in a covariant manner has proven difficult due to the absence of a globally defined time direction. Early approaches to this problem include Komar's covariant formula for conserved charges associated with Killing vectors\cite{Komar:1958wp} and the ADM formalism\cite{Arnowitt:1959ah,Arnowitt:1960es,Arnowitt:1962hi}, which defines charges for asymptotically flat spacetimes. Additionally, Bondi and colleagues developed methods for charges at null infinity\cite{Bondi:1962px,Sachs:1962zza}, and these concepts were extended to asymptotically anti-de Sitter (AdS)\cite{Ashtekar:1984zz,Ashtekar:1999jx,Aros:1999id}, de Sitter (dS) spacetime\cite{Balasubramanian:2001nb}, and higher-derivative gravity theories\cite{Deser:2002jk,Deser:2002rt}. Despite their utility, these methods have limitations, such as lack of full covariance and dependence on specific actions or asymptotic field behavior (see Sec. 4.3 of \cite{Poisson:2009pwt} for a pedagogical introduction to mass and angular momentum in these different forms). A simple, universal formula that works for every asymptotic behavior and every theory seems to not exist in these settings. This highlights the ongoing challenges in defining conserved quantities in GR.

Compared with ADM, Bondi and Komar's methods, the covariant phase space formalism (CPSF) provides a systematic way to obtain conserved charges in generic theories with gauge symmetries, in particular generally covariant theories. This method was initiated in \cite{Lee:1990nz,Henneaux:1992ig,Ashtekar:1987hia,Witten:1986qs,Crnkovic:1986ex,Crnkovic:1987tz} and studied further in \cite{Barnich:2007bf,Wald:1993nt,Wald:1999wa,Barnich:2001jy,Szabados:2004xxa,Compere:2007az,Compere:2008us,Harlow:2019yfa,Fiorucci:2021pha,Shi:2020csw,Ciambelli:2022vot}. Wald's approach (or called Wald's formalism) is based on action formulation. Besides, there is another way of formulating the CPSF, given by Barnich et al., based on the equation of motion (EOM)\cite{Barnich:2001jy,Barnich:2007bf}. We shall center on Wald's approach. The crucial object of the CPSF lies in the construction of the phase space and its symplectic structure. For the construction of this phase space, the Lagrangian formulation is required rather than the Hamiltonian one; consequently, its name incorporates ``covariant". After the construction of the phase space, one can deduce surface charges from the symplectic potential. For more details see Sec. \ref{cps} and the literature mentioned above (see \cite{Hollands:2005wt} for a nice comparison of various notions of conserved charges in asymptotically AdS spacetimes).

In 2016, based on the CPSM, Hajian et al. formulated the solution phase space method (SPSM) for computing the conserved charges\cite{Hajian:2015xlp}, such as mass, angular momentum, electric charge, and entropy, which are associated with the exact symmetries of black hole solutions. Distinct from the CPSM, the SPSM enables the calculation of conserved charges via integration over nearly any smooth codimension-2 surface encircling the singularity of the black hole, not being confined to the horizon bifurcation surface. Here, black hole entropy is regarded as a conserved charge connected to a linear combination of generators for stationarity, axial isometry, and other symmetries, with coefficients determined by the selected horizon. This methodology facilitates the derivation of the first law of black hole thermodynamics for any chosen horizon in a black hole solution, addressing the problems identified in previous investigations by circumventing the constraints of specific horizons or asymptotic conditions. 

After the work of Hajian et al., there was some work that applied their methods to other situations\cite{Hajian:2016kxx, Ghodrati:2016vvf, Ding:2020bwa}. However, the circumstances they take into account are relatively complicated. If this method is general, then it must also be applicable to simpler situations. With a simpler case, we can show the details of the method more clearly and know better where we are going. Based on this, we consider the simplest possible scenario, i.e. non-charged rotating BTZ black hole\cite{Banados:1992wn,Banados:1992gq,Carlip:1995qv,Carlip:1994hq,Carlip:1995zj}. In other words, this paper has some pedagogical significance. We strive to provide as many details as possible to show readers how to obtain conserved charges through the SPSM step by step and make everything as reasonable as possible. Some points that need to be noted have also been pointed out. It is hoped that this method will be more widely accepted and applied. Of course, the BTZ black hole itself is also very important. The various problems associated with the semi-classical or quantum aspects of black holes (microstates and entropy, evaporation and unitarity, entanglement, smoothness of internal region, holographic aspects, etc.) arise independently of dimensions. If they can be resolved in a technically more manageable framework, such as in low-dimensional gravity, we can learn valuable lessons about the semi-classical or quantum aspects of dimension-independent black holes. If we want to keep the defining properties of a black hole, an event horizon, then we need at least one dimension of space and one dimension of time. Therefore, the minimum possible size of a black hole toy model is two. But the Einstein tensor vanishes kinematically for any two-dimensional metric, and if we want to make the theory have dynamic effects, we must add some other fields, e.g. scalar field (called dilaton), which corresponds to JT gravity or other models\cite{Mertens:2022irh,Grumiller:2002nm,Blommaert:2018iqz,Grumiller:2021cwg,Blommaert:2018oro,Ruzziconi:2020wrb}. Therefore, the minimum possible dimension of a GR-based black hole toy model is three. On the other hand, the most typical solution of a three-dimensional black hole is the BTZ black hole. For the reasons mentioned above, BTZ black holes and their related three-dimensional gravity have been widely concerned by people, even today. For earlier work on BTZ black holes and three-dimensional gravity, see\cite{Jackiw:1984je,Brown:1988am,Witten:1988hc,Banados:1998gg,Carlip:1998uc,Carlip:2004ba,Witten:2007kt,Myung:2006sq}; For recent work on BTZ black holes and three-dimensional gravity, see\cite{Cotler:2018zff,Geiller:2020edh,Donnay:2016iyk,Cotler:2018zff,Geiller:2020edh,Nguyen:2021pdz,Collier:2023fwi,Upadhyay:2019hyw,Yu:2021rfg,Verheijden:2021yrb,Balthazar:2021xeh,Hennigar:2020drx,Ditta:2022fjz,Craps:2020ahu,Narzilloev:2021jtg,Du:2023nkr,Craps:2021bmz,EslamPanah:2022ihg,Bueley:2022ple,Santos:2021orr,Robbins:2021ion,Jeong:2023rck,Ding:2023niy,Pourhassan:2022cvn,Zeng:2019huf,Pourhassan:2023qpv,Matsueda:2014aoa}. Because BTZ black holes are so important, deepening our understanding of them from any aspect would be beneficial for our community, and we hope our work can make some modest contributions to this. 

After the work of Hawking et al.\cite{Gibbons:1977mu}, in these years, dS spacetime has attracted a lot of attention\cite{Chandrasekaran:2022cip,Maldacena:2019cbz,Balasubramanian:2020xqf,Geng:2021wcq,Shaghoulian:2022fop,Ovalle:2021jzf,Emparan:2022ijy}. The three-dimensional Kerr-dS spacetime (i.e., KdS$_3$) represents an almost simplest case\cite{Park:1998qk,Wang:2006eb}. Naturally, its thermodynamic properties are a matter of great concern. The computation in KdS$_3$ spacetime using the SPSM is very similar to the case of BTZ black hole. Hence, we also give the results of KdS$_3$ spacetime as a direct extension of the case in the BTZ black hole.

The rest of this paper is organized as follows. In Sec. \ref{solution_phase_space}, we give a review of the solution phase space, including a relatively detailed introduction to the covariant phase space. In Sec. \ref{btz_bh}, we give a review of the BTZ black hole, including a detailed derivation of the surface charge density. In Sec. \ref{charge_1st_law}, we first review Wald entropy; then we obtain conserved charges of the BTZ hole by specific calculations; next, the first law for the BTZ black hole is discussed and the Smarr relation is also obtained incidentally; for completeness, outer horizon and inner horizon cases have been all included; finally, the results of three-dimensional Kerr-dS  spactime have also been obtained. In Sec. \ref{conclusion}, we give our conclusions and some comments.

\section{Solution phase space method}\label{solution_phase_space}
The SPSM is based on the CPSF, so before we begin to describe the SPSM formally, let us go to the CPSF first.

\subsection{Covariant phase space formalism}\label{cps}

The phase space is a manifold $\mathcal{M}$ equipped with a symplectic 2-form $\bm{\Omega}$. In order to obtain $\bm\Omega$ for a generally invariant theory, we consider an arbitrary action
\begin{align}
    S[\Phi]=\int\bm{L},
\end{align}
where $\bm{L}$ is a Lagrangian $d$-form, $\bm{L}=*\mathcal{L}=\bm{\epsilon}\mathcal{L}$ with Lagrangian density $\mathcal{L}$ and volume form $\bm\epsilon$, and $\Phi$ denotes all the dynamical fields. 

Varying the above action, one can obtain

\begin{align}
   \delta\bm{L}[\Phi]=\bm{E}_\Phi\delta\Phi+\dd\bm\Theta(\delta\Phi,\Phi), 
\end{align}
where $\bm{E}_\Phi\delta\Phi$ is the bulk term, in which $\delta\Phi$ is a generic field variation and serves as the basis for the tangent space of the phase space $\mathcal{M}$. The EOM is given by $\bm{E}_\Phi=0$. $\dd\bm\Theta(\delta\Phi,\Phi)$ is the boundary term, in which $\bm\Theta(\delta\Phi,\Phi)$ is a $(d-1)$-form on the spacetime and 1-form on the tangent bundle of $\mathcal{M}$ (or just called field configuration space), i.e., $\bm\Theta$ is a $(d-1;1)$-form, sometimes called as Lee–Wald symplectic potential. When $\bm E_\phi=0$, we have
\begin{align}
    \delta\bm{L}[\Phi]\approx\dd\bm\Theta(\delta\Phi,\Phi),
\end{align}
where ``$\approx$" denotes on-shell equalities.  The
Lee–Wald symplectic form is defined by\cite{Lee:1990nz}
\begin{align}
\bm{\Omega}(\delta_{1}\Phi,\delta_{2}\Phi,\Phi):=\int_{\Sigma}\bm{\omega}(\delta_{1}\Phi,\delta_{2}\Phi,\Phi),
\end{align}
where $\Sigma$ is a codimension-1 Cauchy surface and one can verify that $\bm\Omega$ is independent of the choice of $\Sigma$; $\delta_1\Phi$ and $\delta_2\Phi$ are two arbitrary field perturbations on the tangent space of $\mathcal{M}$; (pre-)symplectic form
\begin{align}
\bm\omega(\delta_1\Phi,\delta_2\Phi,\Phi):=\delta_1\bm\Theta(\delta_2\Phi,\Phi)-\delta_2\bm\Theta(\delta_1\Phi,\Phi),
\end{align}
which is a $(d-1;2)$-form. When restricted to solutions of the EOM, with $\delta_{1,2}\Phi$ satisfying the linearized EOM, one has
\begin{align}\label{dOmega}
    \dd \bm\omega(\delta_1\Phi,\delta_2\Phi,\Phi)\approx 0.
\end{align}
In this sense, we call $\bm\Omega$ is conserved. Denoting the set of solutions to the EOM by $\mathcal{P}$, then $(\mathcal{P};\bm\Omega)$ is the well-defined phase space. The tangent space of $\mathcal{P}$, spanned by $\delta\Phi$, is denoted by $T_{\mathcal{P}}$. In fact, there are still some freedoms in $\bm\omega$, one can give $\bm \Theta$ a shift, i.e. $\bm\Theta(\delta\Phi,\Phi)+\dd \bm \ell(\delta\Phi,\Phi)$, where $\bm \ell$ is a $(d-2;1)$-form corner term, which is not specified by the theory, may be chosen based on physical requirements. One can often assume that this freedom has been fixed. Many literature have touched this problem\cite{Compere:2015bca,Compere:2015mza,Compere:2008us,Compere:2014cna,Compere:2015knw,Freidel:2020xyx,Freidel:2020svx,Freidel:2020ayo}. On the other hand, one can easily check when $\bm\Theta(\delta\Phi,\Phi)\rightarrow\bm\Theta(\delta\Phi,\Phi)+\dd\bm \ell(\delta\Phi,\Phi)$, symplectic form $\bm\omega(\delta_1\Phi,\delta_2\Phi,\Phi)\rightarrow\bm\omega(\delta_1\Phi,\delta_2\Phi,\Phi)+\mathrm{d}[\delta_1\bm \ell(\delta_2\Phi,\Phi)-\delta_2\bm \ell(\delta_1\Phi,\Phi) ]$. In our paper, we just consider the exact symmetries as described below, where $\delta_\eta\Phi=0$. Hence, this freedom does not affect our results and we can ignore it. 


After defining the symplectic form $\bm\omega$, one can further discuss conserved charges corresponding to a specific set of transformations generated by $\delta_\epsilon\Phi$. For this paper, we just consider diffeomorphism transformations, i.e. $\delta_\xi g_{\mu\nu}$ with Killing vector $\xi$. Of course, when one adds more fields, other gauge transformations may need to be considered, for example, the gauge transformation of the electromagnetic field $\delta_\lambda A$. For more explicitness, we will just use $\delta_\xi\Phi$ in the following part. Due to Eq. \eqref{dOmega}, one can construct
\begin{align}
\bm\omega(\delta\Phi,\delta_\xi\Phi)\approx\dd\bm k_\xi(\delta\Phi,\Phi),
\end{align}
where $\bm k_\xi$ is a $(d-2;2)$-form, one can call it (surface) charge density. After integrating $\bm k_\xi$ over a codimension-2 slice, one can define the charge variation associated with $\xi$\cite{Iyer:1994ys}:
\begin{align}\label{charge}
    \slashed{\delta} Q_\xi(\Phi)=\int_\Sigma\mathrm{d}\bm{k}_\xi(\delta\Phi,\Phi)=\oint_{\partial\Sigma}\bm{k}_\xi(\delta\Phi,\Phi).
\end{align}
The charge variation is a $(0;1)$-form and $\partial\Sigma$ is usually a spacelike surface at the boundary of $\Sigma$, people often call it surface charge\cite{Frodden:2019ylc}. The symbol "$\slashed{\delta}$" just reminds us that $\delta(\slashed{\delta}Q_\xi(\Phi))=0$ is not necessary. One can check when $\Sigma$ is chosen as a constant time slice at $t$, then $\slashed{\delta}Q_\xi(\Phi)$ is independent of $t$ or the choice of $\Sigma$. $\slashed{\delta} Q_\xi(\Phi)$ is conserved in the above sense.

When $\slashed{\delta}Q_\xi(\Phi)$ is integratable, i.e. $(\delta_1\delta_2-\delta_2\delta_1)Q_\xi(\Phi)=0$, where $\Phi$ is any field configuration in the phase space $\mathcal{P}$, and $\delta_{1,2}\Phi$ are two elements on its tangent bundle, then the charge $Q_\xi(\Phi)$ is well defined. After that, one can just let $\delta Q_\xi=\slashed{\delta}Q_\xi$, more details see Appx. \ref{IntCon} and references\cite{Hajian:2015xlp,Compere:2015knw}.

If $\slashed{\delta}Q_\xi(\Phi)$ is integrable, and the surface integrals of charge variations are finite and non-vanishing, the charges $Q_\xi(\Phi)$ can be defined. One can easily check using the above equations, we have
\begin{align}
    \bm k_\xi(\delta\Phi,\Phi)\approx\delta\bm Q_\xi-\xi\cdot\bm\Theta(\delta\Phi,\Phi),
\end{align}
where the dot represents the contraction of a vector index and the first index of a differential form, $(d-2;0)$-form $\bm Q_\xi$ is the Noether-Wald charge density, which is related to the Noether current by $\bm J_\xi=\dd\bm Q_\xi$ and $\bm J_\xi$ can also be expressed as
\begin{align}
    \bm J_\xi=\bm\Theta(\delta_\xi\Phi,\Phi)-\xi\cdot\bm L.
\end{align}

So far, we have given all the equipment, and one can obtain the conserved charge $Q_\xi(\Phi)$ from Lagrangian $d$-form $\bm L$ using them. Before fulfilling our previous promises, we should distinguish two symmetries. From Eq. \eqref{charge}, we can know if one wants to guarantee the conservation of charge variation $\slashed{\delta}\bm Q_\xi$ and the independence of integration on $\Sigma$ and $\partial\Sigma$, the symplectic current $\bm\omega$ should vanish on-shell for a subclass of $\delta_\xi\Phi$'s, which can be denoted by $\delta_{\bm\omega}\Phi$, then we have
\begin{align}  \bm\omega(\delta\Phi,\delta_{\bm\omega}\Phi,\Phi)\approx 0
\end{align}
for all $\delta\Phi$ satisfying the linear EOM, the charges can be defined on any compact codimension-2 surface that need not be the boundary of a Cauchy surface. The transformations generated by $\delta_{\bm\omega}\Phi$ are called symplectic symmetries. According to the property of field perturbations, symplectic symmetries can be divided into two sets: 
\begin{itemize}
    \item (Symplectic) non-exact symmetries: these symmetries are a subset of $\delta_{\bm\omega}\Phi$ that are non-vanishing at least at one point in the phase space. When the symmetry generators are denoted by $\chi$, then $\delta_\chi\Phi\neq 0$. These symmetries correspond to symplectic charges\cite{Compere:2015bca,Compere:2015mza,Compere:2015knw}.
    \item (Symplectic) exact symmetries: these symmetries are the subset of $\delta_{\bm\omega}\Phi$ that vanish over the whole phase space. When the symmetry generators are denoted by $\eta$, then $\delta_\eta\Phi= 0$. Exact symmetries do not move us in the solution, so they are the gauge transformation. These symmetries can correspond to Komar-type charges\cite{Hajian:2015xlp}.
\end{itemize}
In this paper, diffeomorphisms with the Killing vectors belong to exact symmetries. Then, the $\bm \ell$-freedom mentioned above has been fixed, and we can ignore it, as previously stated.

\subsection{Solution phase space method}
With the above preparations, let's officially embark on our journey. The SPSM is based on the CPSF. The manifold of the SPSM is a submanifold of $\mathcal{P}$, denoted as $\mathcal{S}$. The tangent space of $\mathcal{S}$ is denoted as $T_{\mathcal{S}}$, which is spanned by non-proper diffeomorphism transformations $\delta_\xi \Phi$ (traditional gauge transformations have been ignored) or by parametric variations $\hat\delta\Phi$. The ``non-proper" means there is a non-trivial surface charge associated with them. Note that diffeomorphism transformations are also exact symmetries as mentioned above. For any theory, there are two different parameters: one is theory parameters, which consists of Newton's constant, the cosmological constant, etc; another is solution parameters, which are typically constants of motion, e.g. mass parameter and angular momentum parameter in this paper, and we just consider these parametric variations. 
 
\begin{figure}
    \centering
    \includegraphics[width=1\linewidth]{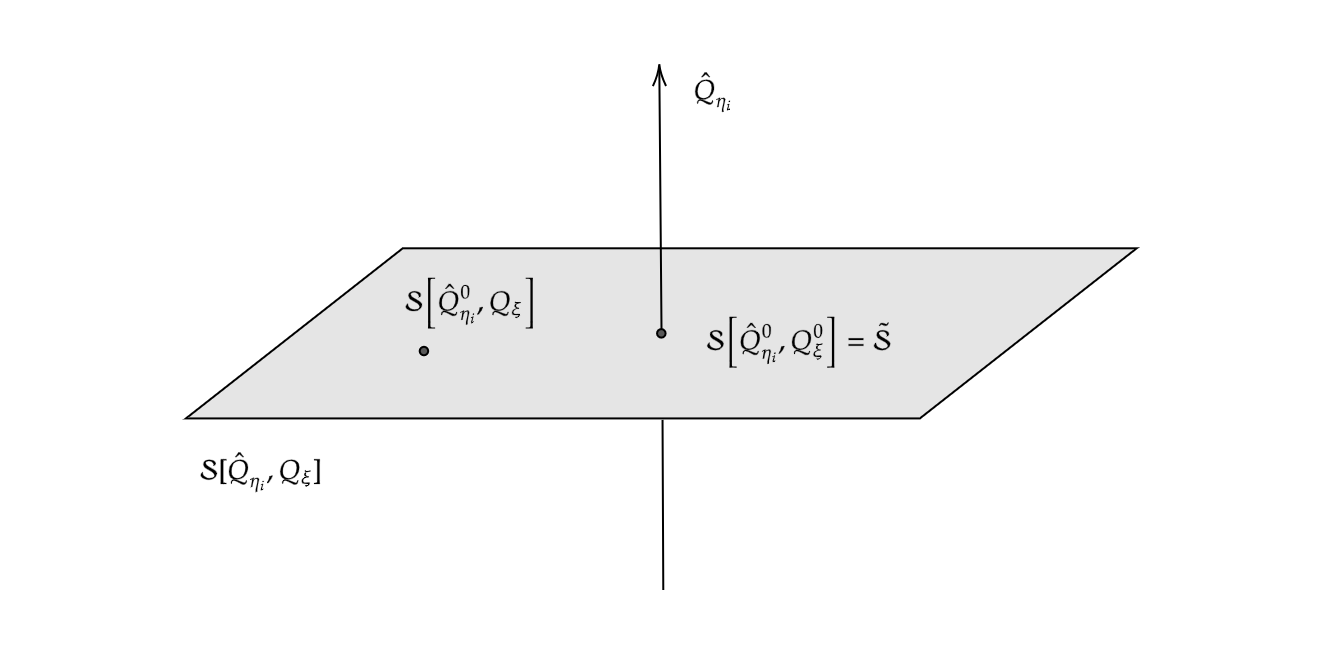}
    \caption{Picture description of solution space $\mathcal{S}[\hat{Q}_{\eta_i},Q_\xi]$. The vertical axis corresponds to different solutions associated with exact symmetry charges $\hat Q_{\eta_i}$ (e.g. mass and angular momentum), i.e., it labels points in $\mathcal{S}_p$. Here, in order to clearly indicate its correlation with parametric variations $\hat{\delta}\Phi$, we have added hats to them. The horizontal plane corresponds to non-proper diffeomorphisms $\xi$, and each point gives us a geometry in the phase space identified by charges $Q_{\xi}$. Moving on the horizontal plane will not change the exact charges $\hat Q_{\eta_i}$, so we think that non-proper exact symmetries are gauge transformations.}
    \label{fig:phase_space}
\end{figure}

Solution parameters are denoted by $p_\alpha$, then parametric variations are 
\begin{align}
    \hat{\delta}\Phi=\pdv{\Phi}{p_\alpha}\delta p_\alpha.
\end{align}
Since non-proper diffeomorphism transformations in this paper are exact symmetries, which do not move us in the solution, we can just focus on the parametric variations. For clarity the above statement, the reader can refer to Fig. \ref{fig:phase_space}.  

$\Phi(x^\mu;p_\alpha)$ with tangent space restricted to $\hat{\delta}\Phi$ further gives us a submanifold $\mathcal{S}_p$ of $\mathcal{S}$, then the corresponding symplectic 2-form can be well defined. 
\begin{align}
\hat{\Omega}:=\Omega(\hat{\delta}_1\Phi,\hat{\delta}_2\Phi,\Phi)=\int_\Sigma\bm{\omega}(\hat{\delta}_1\Phi,\hat{\delta}_2\Phi,\Phi)
\end{align}
with
\begin{align}
    \bm{\omega}(\hat{\delta}_1\Phi,\hat{\delta}_2\Phi,\Phi)=\hat{\delta}_1\bm{\Theta}(\hat{\delta}_2\Phi,\Phi)-\hat{\delta}_2\bm{\Theta}(\hat{\delta}_1\Phi,\Phi) .
\end{align}
Then $(\mathcal{S}_p;\hat{\Omega})$ forms a phase space, which is what we need in this paper. We can further define the charge variation associated with an exact symmetry $\eta$ on $(\mathcal{S}_p;\hat{\Omega})$ as 
\begin{align}
\hat{\slashed\delta}Q_{\eta}=\oint_{\partial\Sigma}\bm{k}_{\eta}(\hat{\delta}\Phi,\Phi) .
\end{align}
Note that we have the parameter-dependent transformation, so $\hat\delta\eta\neq 0$. 

When the integrability condition is satisfied, one can just let $\hat\delta{Q}_\eta=\hat{\slashed\delta}Q_{\eta}$ as mentioned above, more details also see Appx. \ref{IntCon} and references\cite{Hajian:2015xlp,Compere:2015knw}. Then one can integrate $\hat\delta{Q}_\eta$ over $\mathcal{S}_p$ to obtain $Q_\eta[\Phi(p_\alpha)]$. To do this, one should choose the appropriate reference point $Q_\eta[\Tilde{\Phi}]$: 
\begin{align}
    Q_{\eta}[\Phi(p_{\alpha})]=\int\limits_{\Tilde{p}}^{p}\hat{\delta}Q_{\eta}+Q_{\eta}[\Tilde{\Phi}] ,
\end{align}
with the integration being on arbitrary integral curves, which connect $\Tilde{\Phi}$ to $\Phi$ on $\mathcal{S}_p$, integrability guarantees that the charge does not depend on the integration path. In this paper, we will choose $Q_\eta[\Tilde{\Phi}(m=j=0)]=0$ as the reference point, which corresponds to the vacuum situation. 

\section{Review of BTZ black hole}\label{btz_bh}
In this section, we will give a short review of the BTZ black hole.

\subsection{Surface charge} 
We just consider a non-charged rotating BTZ black hole, where the dynamical fields $\Phi$ are just the metric field $g_{\mu\nu}$, so the action is
\begin{align}
    S[g_{\mu\nu}]=\int\bm L,
\end{align}
where $\bm L$ is a Lagrangian $3$-form, i.e. 
\begin{align}
    \bm{L}=\bm{\epsilon}\mathcal{L}=\frac{\sqrt{-g}}{6}\epsilon_{\mu\nu\rho}\mathcal L\mathrm{~d}x^{\mu}\wedge\mathrm{d}x^{\nu}\wedge\mathrm{d}x^{\rho},
\end{align}
in which $\mathcal{L}$ is the customary Einstein-Hilbert bulk Lagrangian density with negative cosmological constant $\Lambda$ ($=-l^{-2}$),
\begin{align}
\mathcal{L}=\frac{1}{2\pi }\Big(R-2\Lambda\Big),
\end{align}
where $R$ is the Ricci scalar; we have let $8G=1$.  

Vary Lagrangian $3$-form, we have 
\begin{align}
    \delta\bm{L}=\bm \epsilon E^{\mu\nu}\delta g_{\mu\nu}+\dd\bm{\Theta}(\delta g_{\mu\nu},g_{\mu\nu}),
\end{align}
where $E^{\mu\nu}=\frac{1}{2\pi}(-R^{\mu\nu}+\frac{1}{2}g^{\mu\nu}R-g^{\mu\nu}\Lambda)$, $E^{\mu\nu}=0$ is the EOM , $\bm{\Theta}(\delta g_{\mu\nu},g_{\mu\nu})$ is the surface (2; 1)-form, 
\begin{align}
    \bm{\Theta}(\delta g_{\mu\nu},g_{\mu\nu})=\frac{\sqrt{-g}}{2}\epsilon_{\mu\nu\rho}\Theta^{\mu}\dd x^{\nu}\wedge\mathrm{d}x^{\rho},
\end{align}
in which 
\begin{align}
    \Theta^{\mu}=\frac{1}{2\pi }\left(\nabla_\nu h^{\mu\nu}-\nabla^\mu h \right)
\end{align}
with $h^{\mu\nu}\equiv g^{\mu\rho}g^{\nu\sigma}\delta g_{\rho\sigma}$, $h\equiv h^\mu\,_\mu$. Then we can obtain the Noether current, 
\begin{align}
    J^\mu=\Theta^{\mu}-\xi^{\mu}\mathcal{L}\approx\frac{1}{2\pi }\nabla_{\nu}\left(\nabla^{\nu}{\xi^{\mu}}-\nabla^{\mu}{\xi^{\nu}}\right).
\end{align}
Whence the $(1;0)$-form Noether-Wald charge density can be obtained by $\bm J_\xi=\dd\bm Q_\xi$,
\begin{align}
    \bm{Q}_{\xi}=\sqrt{-g}\epsilon_{\mu\nu\rho}Q_{\xi}^{\mu\nu}\mathrm{d}x^{{\rho}},
\end{align}
where 
\begin{align}
    Q_{\xi}^{\mu\nu}\approx\frac{1}{4\pi }(\nabla^{\nu}\xi^{\mu}-\nabla^{\mu}\xi^{\nu}).
\end{align}
More details of derivation can be found in the Appx. \ref{app:charge}. Finally, we can find the (1; 1)-form surface charge density, 
\begin{align}
    \bm{k}_{\xi}(\delta g_{\mu\nu},g_{\mu\nu})\approx\delta\bm{Q}_{\xi}-\xi\cdot\bm{\Theta} 
\end{align}
with 
\begin{align}
   \bm{k}_{\xi}(\delta g_{\mu\nu},g_{\mu\nu}) =\sqrt{-g}\epsilon_{\mu\nu\rho}k_{\xi}^{\mu\nu}\mathrm{d}x^{\rho},
\end{align}
where
\begin{widetext}
 \begin{align}\label{charge_density}
k_{\xi}^{\mu\nu}\approx\frac{1}{2\pi }(\xi^{[\mu}\nabla_{\sigma} h^{\nu]\sigma}-\xi^{[\mu}\nabla^{\nu]} h-\xi_{\sigma}\nabla^{[\mu} h^{\nu]\sigma}-\frac{1}{2} h\nabla^{[\mu}\xi^{\nu]}+h^{\sigma[\mu}\nabla_{\sigma}\xi^{\nu]}).
\end{align}
\end{widetext}
Integrating it over a codimension-2 surface $\partial\Sigma$, we get a charge variation (or call surface charge), 
\begin{align}\label{surface_charge}
\slashed{\delta}Q_\xi(\delta g_{\mu\nu},g_{\mu\nu})=\oint_{\partial\Sigma}\bm{k}_\xi(\delta g_{\mu\nu},g_{\mu\nu}).
\end{align}
When the integrability condition is satisfied, we have $\slashed{\delta} Q_\xi=\delta Q_\xi$ (more details see Appx. \ref{IntCon}), which is independent of the choice of integration surface $\partial\Sigma$ surrounding the singularity of the BTZ black hole.

\subsection{Physics of BTZ black hole}
Let's talk about some physics of the BTZ black hole, i.e., give some setups for the next discussion. In this paper, we just talk about a non-charged rotating BTZ black hole in Schwarzschild-like coordinates, which is described by the following metric ansatz
\begin{align}\label{metric}
\dd s^{2}=-N^{2}\dd t^{2}+N^{-2}\dd r^{2}+r^{2}(N_{\phi}\dd t+\dd\phi)^{2} 
\end{align}
with lapse and shift functions
\begin{align}
N(r)=\sqrt{-m+\frac{r^{2}}{l^{2}}+\frac{j^{2}}{4r^{2}}},\quad  N_{\phi}(r)=-\frac{j}{2r^{2}},
\end{align}
where $-\infty<t<\infty$, $0<r<\infty$ and $0<\phi<2\pi$; $m$ is the mass parameter and $j$ is the angular momentum parameter, which are solution parameters we need to consider as mentioned above. In the horizon, the lapse function $N(r)$ vanishes, which gives 
\begin{align}
r_\pm=\sqrt{\frac{l \left(l m\pm\sqrt{l^2 m^2-j^2}\right)}{2}},
\end{align}
which are two coordinate singularities, where $r_+$ corresponds to the black-hole outer horizon, while $r_-$ corresponds to the black-hole inner horizon. When $r_-=r_+$, this is the case for an extremal black hole, where $|j|=ml$\footnote{In the extremal BTZ black hole, $r_0=r_+=r_-$, the metric becomes $\dd s^2=-\frac{(r^2-r_0^2)^2}{l^2 r^2}\dd t^2+\frac{l^2 r^2}{(r^2-r_0^2)^2}\dd r^2+r^2(\dd\phi-\frac{r_0^2}{lr^2}\dd t)^2$. For the extremal black holes, there is no bifurcation surface, so the idea of deriving entropy as a Noether charge does not work. However, from a physical perspective, one can expect entropy to be a continuous function for near-extremal black holes. Therefore, the entropy of extremal black holes can be obtained by a limiting procedure starting from near-extremal black holes, and the NHEG phase space and mechanics may need to be considered\cite{Compere:2015mza,Compere:2015bca,Hajian:2013lna}. In the SPSM, the entropy need not be computed by an integration
over the codimension-2 bifurcation surface of the bifurcate
Killing horizon, and it could be computed by integrating the
corresponding density over any closed, compact spacelike
codimension-2 surface. Therefore, the SPSM can still be used in the extremal case, a specific example see \cite{Hajian:2015xlp}, where the integrability is also discussed. More information for  the (near-)extremal black hole, see also \cite{Kunduri:2007vf,Compere:2015mza,Balasubramanian:2009bg,Gupta:2008ki,Hajian:2015eha,Seraj:2016cym,Kunduri:2013gce,Johnstone:2013ioa}.}. We can also get
\begin{align}
m=\frac{r_+^2+r_-^2}{l^2},\quad j=\frac{2r_+ r_-}{l}.
\end{align}
The time-time component $g_{00}$ of the metric vanishes at $r_{erg}=m^{1/2}l=\sqrt{\rp^2+\rm^2}$. $r_{erg}$ is the surface of infinite redshift, and we have $r_-\leq r_+\leq r_{erg}$. The region between $r_+$ and $r_{erg}$ is the ergosphere. When $j > 0$, all observers in this region will be dragged along by the rotation of the black hole.

In order to describe a black hole, one must have $m>0$ and $|j|\leq ml$. When $|j|>ml$, the horizons disappear, leaving a metric that has a naked conical singularity at $r=0$. In particular, when $m=-1,j=0$, the metric just corresponds to AdS$_3$ spacetime, where the singularity disappears. There is no horizon, and there is also no singularity to hide. When $m=j=0$, one turns to the massless black hole.

The metric \eqref{metric} is stationary and axially symmetric, so there are only two Killing vectors $\partial_t$ and $\partial_\phi$, whose linear combination is still a Killing vector. For a stationary black hole, there exists a Killing vector $\xi_H$, which is normal to the horizon of the black hole. We can obtain it by taking a linear combination of $\partial_t$ and $\partial_\phi$, i.e.
\begin{align}
\xi_H=\partial_t+\Omega_H\partial_\phi,
\end{align}
which is called the horizon Killing vector, and the constant $\Omega_H$ is called the angular velocity of the horizon. Since the horizon is a null surface and $\xi_H$ is normal to the horizon, we have $|\xi_H|^2=0$ at the horizon. Hence, we have
\begin{align}
    \Omega_{H_{(\pm)}}= -N_\phi(r_\pm)=\frac{r_\mp}{lr_\pm},
\end{align}
where $H_{(+)}$ denotes the outer horizon, while $H_{(-)}$ denotes the inner horizon.

In fact, one can check that the Killing vector $\xi_H$ does not satisfy the integrability condition, but after redefining $\zeta_H=\frac{2\pi}{\kappa}\xi_H$, where $\kappa$ is the surface gravity\cite{Wald:1984rg},
\begin{align}
 \kappa_{(\pm)}=\sqrt{-\frac{1}{2}(\nabla_{\mu}\xi_{\nu})^{2}\Big|_{H_{(\pm)}}}=\pm\frac{r_+^2-r_-^2}{l^2r_\pm},
\end{align}
in which the negative sign of $\kappa_{(-)}$ comes from thermal instability of the inner horizon and is related to the cosmic censorship conjecture\cite{Yurtsever:1992th}. The integrability condition can be satisfied (see Appx. \ref{IntCon}). Next, when we discuss black hole entropy, we just use this redefined Killing vector $\zeta_H$.

\section{Conserved charges and first laws for BTZ black holes}\label{charge_1st_law}
\subsection{Black hole entropy}
For completeness, we will give some introductions about the Wald entropy. In Sec. \ref{cps}, we have given the introduction of the CPSM. In the 1990s, Wald and his collaborators made the proposition based on the CPSM, where the black hole entropy can also be treated as a Noether charge\cite{Wald:1993nt,Iyer:1994ys}.

The definition of the Wald entropy is 
\begin{align}
S_{BH}^W:=2\pi\int_{\partial\Sigma}\bm{X}^{\mu\nu}\bm{\epsilon}_{\mu\nu},
\end{align}
where $\bm{\epsilon}_{\mu\nu}$ is the binormal tensor to the $(d-2)$-bifurcation surface $\partial\Sigma$, normalized as $\bm{\epsilon}_{\mu\nu}\bm{\epsilon}^{\mu\nu}=-2$, and $\bm{X}^{\mu\nu}$ is defined by Noether-Wald charge
\begin{align}
     \bm{Q}_\xi=\bm{X}^{\mu\nu}(\Phi)\nabla_{[\mu}\xi_{\nu]}+\bm{W}_\mu(\Phi)\xi^\mu+\bm{Y}(\Phi,\delta_\xi\Phi)+\mathrm{d}\bm{Z}(\Phi,\xi),
\end{align}
where $\bm X^{\mu\nu}, \bm W_\mu,\bm Y, \bm Z$ are covariant quantities that are locally constructed from the dynamic fields and their derivatives, $\bm Y$ is linear in $\delta_\xi\Phi$, and $\bm Z$ is linear in $\xi$. Note that $\bm X^{\mu\nu}, \bm W_\mu,\bm Y, \bm Z$ have many different choices (also called $\bm X-, \bm W-,\bm Y-, \bm Z-$ ambiguities) and $\bm Q_\xi$ is not unique. We can choose 
\begin{align}
    (\bm X^{\mu\nu})_{\mu_3\cdots\mu_d}=-\pdv{\mathcal{L}}{R_{\alpha\beta\mu\nu}}\bm\epsilon_{\alpha\beta\mu_3\cdots\mu_d},
\end{align}
where $R_{\alpha\beta\mu\nu}$ and $\bm\epsilon$ are the Riemann tensor and volume form, respectively. When considering the case on the horizon, due to $\xi_{H}$ being a Killing vector and $\partial\Sigma$ being a bifurcation surface, the $\bm W-,\bm Y-, \bm Z-$ ambiguities can be removed. Using
\begin{align}
(\dd\xi_{H})_{\mu\nu}=2\kappa\bm{\epsilon}_{\mu\nu},
\end{align}
we have 
\begin{align}
Q_{\zeta_{H}}=\int_{\partial\Sigma}\bm{Q}_{\zeta_{H}}=S_{BH}^{W},\quad\zeta_{H}:=\frac{2\pi}{\kappa}\xi_{H}.
\end{align}
We can conclude that entropy is the Noether charge associated with the Killing vector $\zeta_H$. Due to the above relation, we can also use the SPSM to compute the black hole entropy. We can define the entropy variation as\footnote{To clarify its rationality, just to note that the variation of the Hamiltonian generator associated with
flows of $\xi_H$ is given by $\delta H_{\xi_H}=\int_H\bm{k}_{\xi_H}(\delta\phi,\phi)$. To satisfy a consistent first law, then the entropy
variation should be defined by $\delta H_{\xi_H}:=T_{H}\delta S$. Hence, we can define the entropy variation as $\delta S:= \frac{1}{T_{H}}\int_H\bm{k}_{\xi_H}(\delta\phi,\phi)$, where $T_H=\frac{\kappa}{2\pi}$, more details see \cite{Iyer:1994ys,Hajian:2015xlp,Hajian:2020dcq}.}
\begin{align}
\hat{\delta}S:=\oint_{\partial\Sigma}\bm{k}_{\zeta_H}(\hat{\delta}\Phi,\Phi)=\frac{2\pi}{\kappa}\oint_{\partial\Sigma}\bm{k}_{\xi_H}(\hat{\delta}\Phi,\Phi),
\end{align}
where $\bm{k}_{\xi_H}$ is just the surface charge density for $\xi_H$.

So far, we have not only demonstrated the rationality of the redefined horizon Killing vector $\zeta_H=\frac{2\pi}{\kappa}\xi_H$ related to black hole entropy from the perspective of integrability (see Appx. \ref{IntCon}), but also from the perspective of Wald charge.

\subsection{Conserved charges}
Next, we will discuss the specific calculations for conserved charges of the BTZ black hole. For the BTZ black hole, there are two Killing vectors, $\partial_t$ and $\partial_\phi$, which correspond to time translation symmetry and spatial rotational symmetry, respectively. By Noether's theorem, we can know that the mass $M$ and angular momentum $J$ of the BTZ black hole correspond to $\partial_t$ and $\partial_\phi$, respectively. From the above discussion, we have known that $\zeta_H=\frac{2\pi}{\kappa}\xi_H$ corresponds to the black hole entropy $S$. So we only need to calculate the surface charges corresponding to $\partial_t,\partial_\phi,\zeta_H$ separately, and then we can obtain $M, J, S$.

In our case, the solution parameters are $m$ and $j$, so the parametric variations of the metric fields are
\begin{align}\label{para_var}
\hat{\delta}g_{\mu\nu}=\pdv{g_{\mu\nu}}{m}\delta m+\pdv{g_{\mu\nu}}{j}\delta j.
\end{align}
Substituting this into Eq. \eqref{charge_density}, we can obtain the surface charge density. But before jumping into it, let's first look at where we want to go, it might make life easier. From Eq. \eqref{surface_charge}, we have the following surface charge
\begin{align}
{\delta}Q_{\xi}(g_{\mu\nu})=\oint_{\partial\Sigma}\bm{k}_{\xi}(\hat\delta g_{\mu\nu},g_{\mu\nu}),
\end{align}
which is independent of the choice of integration surface $\partial\Sigma$, so we can take $\partial\Sigma$ to be the circle of constant $(t,r)$ and further take the limit $r\rightarrow\infty$ for simplicity\footnote{This approach to the limit is similar to the situation of the quasilocal energy and angular momentum of Brown and York\cite{Carlip:1995zj,Brown:1992br,Brown:1994gs,Mann:1995eu}.}. Then, we have 
\begin{align}\label{surface_charge_reduced}
\delta Q_\xi(\delta g_{\alpha\beta},g_{\alpha\beta})=2\lim_{r\rightarrow\infty}\int_0^{2\pi}\sqrt{h}k_\xi^{tr}\dd\phi=4\pi\lim_{r\rightarrow \infty}  r k_\xi^{tr}.
\end{align}
Hence, we just need to compute $t,r$-part for $k^{\mu\nu}_\xi$. Substituting Eq. \eqref{para_var} into Eq. \eqref{charge_density}, we can obtain 
\begin{align}
k_{\partial_t}^{tr}=&-k_{\partial_t}^{rt}=\frac{\delta m}{4\pi  r},\\ 
     k_{\partial_\phi}^{tr}=&-k_{\partial_\phi}^{rt}=-\frac{\delta j}{4\pi  r},\\
      k_{\zeta_H}^{tr}=&-k_{\zeta_H}^{rt}=\frac{ \delta m- \Omega_H\delta j }{2 \kappa  r}.
\end{align}
The complete results see Appx. \ref{app:charge}. Substituting it to Eq. \eqref{surface_charge_reduced}, we have
\begin{align}
{\delta}Q_{\partial_t}=&\delta m,\\ 
    {\delta}Q_{\partial_\phi}=&-\delta j,\\ 
    {\delta}Q_{\zeta_H}=&\frac{2 \pi(\delta m- \Omega_H\delta j) }{\kappa  }.
\end{align}
Integrating them, we can obtain
\begin{align}
    Q_{\partial_t}=&m,\\ 
    Q_{\partial_\phi}=&-j,\\ 
    Q_{\zeta_{H_{(\pm)}}}=&4\pi r_\pm,
\end{align}
where we have chosen the reference points for the charges at $Q_{\xi}=0$ with $m=j=0$ to fix the integration constant. 

Finally, we can obtain 
\begin{align}
    M:=&Q_{\partial_t}=m,\\ 
    J:=&-Q_{\partial_\phi}=j,\\ 
    S_{(\pm)}:=&Q_{\zeta_{H_{(\pm)}}}=4\pi r_\pm=2L_\pm,
\end{align}
where $L_+$ and $L_-$ are the perimeters of the outer and inner Killing horizons, respectively. These results are consistent with Refs. \cite{Carlip:2005zn,Carlip:1995qv,Banados:1992gq,Carlip:1994hq,Hajian:2015eha,Hajian:2021hje}.

\subsection{First law of black hole thermodynamics}
From the zero law of black hole thermodynamics, i.e. $T_H=\frac{\kappa}{2\pi}$, we can obtain the temperature of the black hole
\begin{align}
    T_{H_{(\pm)}}=\pm\frac{r_+^2-r_-^2}{2\pi l^2 r_\pm}.
\end{align}
Combining results in the last section, we can find
\begin{align}
\delta M=T_H\delta S+\Omega_H\delta J.
\end{align}
This is nothing but the first law of black hole thermodynamics for our case. 

We can also obtain the Smarr relation\cite{Smarr:1972kt,Wang:2006eb} incidentally\footnote{For asymptotic flat black holes in any dimenstions, the general Smarr relation is $\frac{D-3}{D-2}M=TS+\Omega J$, where $D$ is the dimension of spacetime\cite{Gauntlett:1998fz}. It is not difficult to discover the Smarr relation we have obtianed cannot fit the general result in the $D=3$ case, which is related to the effect of the cosmological constant, more details see \cite{Wang:2006eb,Caldarelli:1999xj,Sekiwa:2006qj,Zeng:2019huf}.}:
\begin{align}
M=&\frac{1}{2}T_{H} S+\Omega_{H} J,
\end{align}
where the results of the inner and outer horizons are the same.  Note that the Smarr type relations are not universal, unlike the first law; their form depends on the theory and on the asymptotic behavior of the solution.

\subsection{Extension: Kerr-dS spacetime}

As we said in the introduction, the case of the KdS$_3$ spacetime  is very similar to the case of BTZ black hole. For completeness, we would give the results of KdS$_3$ spacetime in this section. For brevity,  we only give the key parts related to the SPSM, more details about KdS$_3$ spacetime see \cite{Park:1998qk,Wang:2006eb}.

As is known, $\Lambda=-l^{-2}$. In BTZ black hole, $\Lambda <0$, while in KdS$_3$ spacetime, $\Lambda >0$. Thus, we can perform a transformation by renaming $l \rightarrow il$ to transition from BTZ to KdS$_3$. After that, we can find that for $j=0$, there is no horizon event when $m>0$. In order to restore the horizon event for $m>0$, we further rename $m\rightarrow-m$. In summary, we can carry out the renaming $l\rightarrow il$ and $m\rightarrow -m$ to transform to the KdS$_3$ spacetime.

Solve $N(r)=0$, we have
\begin{align}
   r_{\pm}= \sqrt{\frac{l(l m\pm  \sqrt{l^2 m^2+j^2})}{2}}
\end{align}
where $r_+$ is the radius of the cosmological horizon, $H$, instead of the radius of a black-hole horizon in the KdS$_3$ spacetime. Since $r_-$ is a pure-imaginary number, we can define 
\begin{align}
    r_{(-)}\equiv -ir_-=\sqrt{\frac{l( \sqrt{l^2 m^2+j^2}-lm)}{2}}
\end{align}
to obtain the corresponding real number $r_{(-)}$. Then, one can easily obtain 
\begin{align}
    m=\frac{r_+^2-r_{(-)}^2}{l^2},\quad j= \frac{2 r_+ r_{(-)}}{l}.
\end{align}
The angular velocity and surface gravity can be evaluated as before
\begin{align}
    \Omega_+=&-N_\phi(r_+)=\frac{r_{(-)}}{lr_+},\\ 
    \kappa=&-\sqrt{-\frac{1}{2}(\nabla_\mu\xi_\nu)^2\Big|_{H}}=-\frac{r_+^2+r_{(-)}^2}{l^2 r_+}.
\end{align}
which are geometry quantities. Similar to the preceding steps, the following surface charge density can be obtained
\begin{align}
{\delta}Q_{\partial_t}=&-\delta m,\\ 
    {\delta}Q_{\partial_\phi}=&-\delta j,\\ 
    {\delta}Q_{\zeta_H}=&-\frac{2 \pi(\delta m+ \Omega_+\delta j) }{\kappa  }.
\end{align}
By integrating them, we can obtain the surface charge as before
\begin{align}
    Q_{\partial_t}=&-m,\\ 
    Q_{\partial_\phi}=&-j,\\ 
    Q_{\zeta_{H}}=&4\pi r_+.
\end{align}
Then we have
\begin{align}
M:=&-Q_{\partial_t}=m, \\ 
J:=&-Q_{\partial_\phi}=j,\\
S:=&4\pi r_+,
\end{align}
which correspond to the mass, angular momentum and entropy of KdS$_3$ spacetime, respectively. The Hawking temperature
and angular velocity can be evaluated,
\begin{align}
    T_H=&-\frac{\kappa}{2\pi}=\frac{r_+^2+r_{(-)}^2}{2\pi l^2 r_+},\\
    \Omega_H=&-\Omega_+=-\frac{r_{(-)}}{lr_+}.
\end{align}
Note that both of them have a minus sign assigned to them in relation to their geometric definitions\cite{Wang:2006eb}. 

Finally, we can obtain the first law of thermodynamics and the Smarr relation for KdS$_3$ spacetime,
\begin{align}
    \delta M=&T_H\delta S+\Omega_H\delta J,\\
    M=&\frac{1}{2}T_{H} S+\Omega_{H} J,
\end{align}
which are same as the results of the BTZ black hole. These results are consistent with Refs. \cite{Wang:2006eb,Park:1998qk,Balasubramanian:2001nb}.

\section{Conclusions and comments}\label{conclusion}
In this paper, we discussed how to use the solution phase space method to obtain the conserved charges of BTZ black holes. Subsequently, the first law of thermodynamics was derived by leveraging the previously obtained conserved charge, and the Smarr relation was incidentally obtained. For completeness, outer horizon and inner horizon cases have been all included. Finally, the results of three-dimensional Kerr-dS  spactime have also been obtained. As the ending, we give some necessary comments. The solution phase space method for computing charges proves to be highly effective in asymptotic AdS scenarios. Utilizing alternative methods (e.g., see \cite{Hollands:2005wt} for a review) to calculate similar charges might result in infinite values that require regularization. The solution phase space method works for any spacetime that has Killing vectors, regardless of its asymptotic behavior. Such a spacetime does not necessarily need to be a black hole, as shown in our calculation of Kerr-dS spacetime. Another example, in near-horizon geometries, this method functions effectively (e.g., see \cite{Hajian:2015xlp} for extremal Kerr-Newman near horizon geometry); however, the ADM formalism or its extensions and the Komar integral method all prove unsuccessful. In the solution phase space method, one does not need to take the integration surface $\partial\Sigma$ to the bifurcation surface of the horizon or $\bm i^0$ at infinity, unlike in \cite{Iyer:1994ys}. The solution phase space method can also be used for the inner horizon and there exists an inner horizon first law \cite{Castro:2012av} as shown in our calculations. As the most direct generalization, one can consider the charged BTZ black hole\cite{Banados:1992wn}, where the gauge symmetry of the electromagnetic field should be considered. For this case, readers can refer to \cite{Ghodrati:2016vvf}. Besides, one can also consider adding more other fields, e.g., dilaton\cite{Chan:1994qa}. In principle, the new surface charges would be added\cite{Frodden:2019ylc}. One can apply the solution phase space method to these cases. It is well known that three-dimensional gravity has a gauge formulation of the Chern-Simons type \cite{Achucarro:1986uwr, Witten:1988hc}. Thus, one can extend our calculation to this scenario and contemplate how our derivation is associated with this case in this form. In recent years, BTZ-like black holes have attracted people's attention\cite{Ditta:2022fjz,Ding:2023niy}, and theoretically, one can also try to extend our method to this situation. Naturally, other variants of BTZ black holes can also be taken into account\cite{Hennigar:2020drx,Narzilloev:2021jtg,EslamPanah:2022ihg,Pourhassan:2023qpv}. As we mentioned above, the (near-)extremal case has unique features\cite{Compere:2015mza,Compere:2015bca,Hajian:2013lna,Kunduri:2007vf,Compere:2015mza,Balasubramanian:2009bg,Gupta:2008ki,Kunduri:2013gce,Johnstone:2013ioa,Craps:2020ahu}, so extending our discussion to this case would be interesting. In recent years, the effect of the cosmological constant has attracted people's attention again\cite{Hajian:2023bhq,Xiao:2023lap,Chernyavsky:2017xwm,Zeng:2019huf}, so further investigation of this situation would be also an interesting research direction. 
Due to the universality of the solution space method, in principle, one is capable of investigating all types of spacetime that satisfy Killing symmetry through it as mentioned above. In these years, questions related to holography and black hole information on the BTZ black hole or three-dimensional Kerr-dS spacetime background have attracted people's attention\cite{Nguyen:2021pdz,Yu:2021rfg,Verheijden:2021yrb,Balthazar:2021xeh,Craps:2021bmz,Bueley:2022ple,Santos:2021orr,Myung:2006sq,Matsueda:2014aoa,Balasubramanian:2001nb}. Considering how to apply the solution phase space method to these issues is also a worthwhile question.

\begin{acknowledgments}
This work is partly supported by the National Key Research and Development Program of China with Grant No. 2021YFC2203001 as well as the National Natural Science Foundation of China with Grants No. 12075026 and No. 12361141825. W.G thanks Kamal Hajian for classifying the case of (near-)extremal black holes. W.G also thanks Kai Shi and Kang Liu for their helpful discussions. W. G also thanks Professor Hongbao Zhang for his great support, trust, and help.

\end{acknowledgments}

\onecolumngrid
\appendix
\section{$p$-form}
For the reference of readers and to provide our symbol conventions, we will give a short introduction to $p$-forms (or just called differential forms). For a more complete introduction, see Refs\cite{Wald:1984rg,Nakahara:2003nw}. For brevity, we will focus on the Lorentz spacetime.

Antisymmetric tensors with $p$ lower indices are also known as $p$-forms. Without indices, we can represent them as
\begin{align}
    \bm\Omega_p:=\frac1{p!} \Omega_{\mu_1\mu_2...\mu_p} \mathrm{d}x^{\mu_1}\wedge\mathrm{d}x^{\mu_2}\wedge\cdots\wedge\mathrm{d}x^{\mu_p},
\end{align}
where ``$\wedge$'' is the wedge product, defined by 
\begin{align}
    \dd x^1\wedge\dd x^2:=\dd x^1\otimes\dd x^2-\dd x^2\otimes\dd x^1,
\end{align}
which is antisymmetric. For $D$-dimensional spacetime, 
\begin{align}
\Omega_{\mu_1\mu_2\cdots\mu_p}=\sqrt{-g}\epsilon_{\mu_1\mu_2\cdots\mu_{D}}\Omega^{\mu_{p+1}\mu_{p+2}\cdots\mu_{D}},
\end{align}
where $\epsilon$ is the tensor density defined by $\tilde{\epsilon}/\sqrt{-g}$ with $\tilde{\epsilon}_{12\cdots D}=+1$, and 
\begin{align}
    \epsilon^{\mu_{1}\cdots \mu_{j}\mu_{j+1}\cdots \mu_{D}}\epsilon_{\mu_{1}\cdots \mu_{j}\nu_{j+1}\cdots \nu_{D}}=\frac{1}{g}(D-j)! j! \delta^{[\mu_{j+1}}_{\nu_{j+1}}\cdots\delta^{\mu_{D}]}_{\nu_{D}}.
\end{align}
Consider the wedge product between a $p$-form and a $q$-form, one can obtain a $(p+q)$-form. For $D$-dimensional spacetime, we have $p\leq D$ due to total antisymmetry. The top-form with $p=D$ is also called the volume form,
\begin{align}
\bm\epsilon_{(D)}=\sqrt{-g} \frac{1}{D!} \epsilon_{\mu_1\mu_2...\mu_D}\dd x^{\mu_1}\wedge\dd x^{\mu_2}\wedge\cdots\wedge \dd x^{\mu_D}.
\end{align}
Cotangent vectors are 1-forms. An important 1-form is the exterior derivative,
\begin{align}
    \bm{\mathrm{d}}=\partial_\mu\dd x^\mu.
\end{align}
People often denote 
\begin{align}
    \bm{\mathrm{d}} \wedge\bm\Omega_p\equiv\dd\bm\Omega_p=\frac{1}{p!}\partial_{[\mu}\Omega_{\mu_1\cdots\mu_p]} \dd x^\mu\wedge\mathrm{d}x^{\mu_1}\wedge\cdots\wedge\mathrm{d}x^{\mu_p},
\end{align}
or
\begin{align}
(\dd\bm\Omega_p)_{\mu\mu_1\cdots\mu_p}=(p+1)\partial_{[\mu}\Omega_{\mu_1\cdots\mu_p]},
\end{align}
so we can just use $\dd$ to denote $\bm{\mathrm{d}}$. One can easily check that if  the exterior derivative acts on quantities that are sufficiently smooth, we have $\dd^2=0$. The exterior derivative obeys the Leibnitz rule, i.e.
\begin{align}
\mathrm{d}\big(\bm\Omega_p\wedge\bm\Omega_q\big)=\big(\mathrm{d}\bm\Omega_p\big)\wedge\bm\Omega_q+(-1)^p\bm\Omega_p\wedge\mathrm{d}\bm\Omega_q .
\end{align}
The poincare lemma states that any closed $p$-form (with $p\geq 1$), $\dd\bm\Omega_p=0$, is at least locally exact, $\bm\Omega_p=\dd\tilde{\bm\Omega}_{p-1}$.

The Hodge-star $*$ converts $p$-forms into $(D-p)$-forms, 
\begin{align}
    *\bm\Omega_{p}=\frac{1}{p!(D-p)!}\Omega_{{\nu_{1}\cdots\nu_{p}}}\epsilon_{{\mu_{1}\cdots\mu_{D-p}}}{}^{{\nu_{1}\cdots\nu_{p}}}\mathrm{~d}x^{{\mu_{1}}}\wedge\cdots\wedge\mathrm{d}x^{{\mu_{D-p}}}.
\end{align}
It is involutive, i.e. $**\bm\Omega_p=\pm\bm\Omega_p$.

As a concrete example, we will show how to derive $j^\mu=2\nabla_\nu Q^{\mu\nu}$ from $\dd \bm Q=\bm J$, where $\bm Q$ is $(D-2)$-form. 

Using the above relations, we can get
\begin{align*}
    (D-1)\nabla_{[\sigma}(Q^{\nu\rho}\epsilon_{|\nu\rho|\mu_1\cdots \mu_{D-2}]})=J^\tau\epsilon_{\tau \sigma \mu_1\cdots \mu_{D-2}}.
\end{align*}
Multiply $\epsilon^{\eta \sigma \mu_1\cdots \mu_{D-2}}$ both sides simultaneously, we have
\begin{align*}
Left=&(D-1)\epsilon^{\eta \sigma \mu_1\cdots \mu_{D-2}}\nabla_{[\sigma}(Q^{\nu\rho}\epsilon_{|\nu\rho|\mu_1\cdots \mu_{D-2}]})
 =(D-1)\epsilon^{\eta [\sigma \mu_1\cdots \mu_{D-2}]}\nabla_{[\sigma}(Q^{\nu\rho}\epsilon_{|\nu\rho|\mu_1\cdots \mu_{D-2}]})\\
   = &(D-1)\epsilon^{\eta \sigma \mu_1\cdots \mu_{D-2}}\epsilon_{\nu\rho \mu_1\cdots \mu_{D-2}}\nabla_{\sigma} Q^{\nu\rho}
      =\frac{1}{g}(D-1) 2!(D-2)!\delta^{[\eta}_\nu\delta^{\sigma]}_\rho\nabla_{\sigma} Q^{\nu\rho}\\
      =&\frac{1}{g}(D-1)! 2!\nabla_{\sigma} Q^{\eta\sigma}\\
Right=&\frac{1}{g}J^\tau (D-1)!\delta_\tau^\eta=\frac{1}{g}J^\eta (D-1)!.
    \end{align*}
Finally, we obtain 
\begin{align}\label{Wald-charge}
    J^{\mu}=2\nabla_\nu Q^{\mu\nu}.
\end{align}

\section{An explicit calculation of the Noether-Wald charge }\label{app:charge}
Here we would like to present an explicit calculation of the Noether-Wald charge. Due to $\bm J=\bm \Theta-\xi\cdot\bm L$, we have
\begin{align*}
    J^{\mu}=&\Theta^\mu-\xi^\mu\mathcal{L}=\frac{1}{2\pi}(\nabla_\nu h^{\mu\nu}-\nabla^\mu h)-\frac{1}{2\pi}\xi^\mu (R-2\Lambda)
    =\frac{1}{2\pi}[\nabla_\nu h^{\mu\nu}-\nabla^\mu h-\xi^\mu(R-2\Lambda)]\\ 
    =&\frac{1}{2\pi}[\nabla_{\nu}(\nabla^\mu\xi^\nu+\nabla^\nu\xi^\mu)-2\nabla^\mu\nabla_\nu\xi^\nu-\xi^\mu(R-2\Lambda) ]
    =\frac{1}{2\pi}[\nabla_{\nu}(\nabla^\mu\xi^\nu+\nabla^\nu\xi^\mu)-2(\nabla_\nu\nabla^\mu\xi^\nu-R^{\mu\nu}\xi_\nu)-\xi^\mu(R-2\Lambda) ]\\ 
    =&
    \frac{1}{2\pi}[\nabla_{\nu}(\nabla^\nu\xi^\mu-\nabla^\mu\xi^\nu)+2\xi_\nu(R^{\mu\nu}-\frac{1}{2}g^{\mu\nu}R+g^{\mu\nu}\Lambda) ]
    \approx\frac{1}{2\pi}\nabla_\nu(\nabla^\nu\xi^\mu-\nabla^\mu\xi^\nu),
\end{align*}
where $\delta_\xi g_{\mu\nu}=\mathscr{L}_\xi g_{\mu\nu}=\nabla_\mu\xi_\nu+\nabla_\nu\xi_\mu$ is used in the forth step, $[\nabla_\alpha,\nabla^\mu]\xi^\alpha=R^{\mu\nu}\xi_\nu$ is used in the fifth step, and the EOM is used in the last step. Using Eq. \eqref{Wald-charge}, we can obtain 
\begin{align}
    Q^{\mu\nu}\approx\frac{1}{4\pi}(\nabla^\nu\xi^\mu-\nabla^\mu\xi^\nu).
\end{align}

\section{Christoffel symbol}
Here we would like to write down the explicit expression of the Christoffel symbol for the BTZ black hole in Schwarzschild-like coordinates. For simplicity, we just list the non-vanishing components.
\begin{align}
\Gamma^t_{\,tr}=&\Gamma^t_{\,rt}=\frac{4 r^3}{j^2 l^2-4 l^2 m r^2+4 r^4}\\
    \Gamma^t_{\,r\phi}=& \Gamma^t_{\,\phi r}=\frac{j r}{2 (-\frac{j^2}{4}-\frac{r^4}{l^2}+m r^2)}\\ 
    \Gamma^r_{\,tt}=&\frac{r (\frac{j^2}{4 r^2}+\frac{r^2}{l^2}-m)}{l^2}\\ 
    \Gamma^r_{\,rr}=&\frac{j^2 l^2-4 r^4}{j^2 l^2 r-4 l^2 m r^3+4 r^5}\\ 
    \Gamma^r_{\,\phi\phi}=&-\frac{j^2}{4 r}-\frac{r^3}{l^2}+m r\\ 
    \Gamma^\phi_{\,tr}=& \Gamma^\phi_{\,rt}=\frac{2 j r}{j^2 l^2-4 l^2 m r^2+4 r^4}\\
    \Gamma^\phi_{\,r\phi}=&\Gamma^\phi_{\,\phi r}=\frac{4 r (r^2-l^2 m)}{j^2 l^2-4 l^2 m r^2+4 r^4}
\end{align}
\section{Surface charge density}\label{app:charge}

Substituting Eq. \eqref{para_var} into Eq. \eqref{charge_density} , we can find the component results of $k^{\mu\nu}_\xi$ for Killing vector $\xi$ for the BTZ black hole:
\newline

\noindent $k_{\partial_t}^{\mu\nu}$: 
\begin{align}
k_{\partial_t}^{tt}=&k_{\partial_t}^{t\phi}=k_{\partial_t}^{\phi t}=k_{\partial_t}^{rr}=k_{\partial_t}^{\phi\phi}=0\\
    k_{\partial_t}^{tr}=&k_{\partial_t}^{rt}=\frac{\delta m}{4\pi r}\\ 
    k_{\partial_t}^{r\phi}=&-k_{\partial_t}^{\phi r}=-\frac{\delta j}{4\pi  l^2 r}
\end{align}
$k_{\partial_\phi}^{\mu\nu}$: 
\begin{align}
k_{\partial_\phi}^{tt}=&k_{\partial_\phi}^{t\phi}=k_{\partial_\phi}^{\phi t}=k_{\partial_\phi}^{rr}=k_{\partial_\phi}^{\phi\phi}=0\\
    k_{\partial_\phi}^{tr}=&-k_{\partial_\phi}^{rt}=-\frac{\delta j}{4\pi  r}\\ 
    k_{\partial_\phi}^{r\phi}=&-k_{\partial_\phi}^{\phi r}=\frac{\delta m}{4\pi  r}
\end{align}
$k_{\zeta_H}^{\mu\nu}$: 
\begin{align}
k_{\zeta_H}^{tt}=&k_{\zeta_H}^{t\phi}=k_{\zeta_H}^{\phi t}=k_{\zeta_H}^{rr}=k_{\zeta_H}^{\phi\phi}=0\\
    k_{\zeta_H}^{tr}=&-k_{\zeta_H}^{rt}=\frac{ \delta m- \Omega_H\delta j }{2 \kappa  r}\\ 
    k_{\zeta_H}^{r\phi}=&-k_{\zeta_H}^{\phi r}=\frac{ l^2 \Omega_H\delta m-\delta j }{2 \kappa  l^2 r}
\end{align}

One can simply substitute $m \rightarrow -m$, $l \rightarrow il$, and then the results for KdS$_{3}$ spacetime can be obtained. For the sake of brevity, the results of this case would not be listed.
\section{Integrability condition}\label{IntCon}
Following Ref. \cite{Hajian:2015xlp}, we know that integrability condition, $(\delta_1\delta_2 -\delta_2\delta_1)Q_\xi(\Phi)=0$, is equivalent to 
\begin{align}
\oint_{\partial\Sigma}\left(\xi\cdot\bm{\omega}(\hat{\delta}_{1}\Phi,\hat{\delta}_{2}\Phi,\Phi)+\bm{k}_{\hat{\delta}_{1}\eta}(\hat{\delta}_{2}\Phi,\Phi)-\bm{k}_{\hat{\delta}_{2}\eta}(\hat{\delta}_{1}\Phi,\Phi)\right)\approx 0.
\end{align}
If $\xi$ is a linear combination of parameter-independent Killing vectors, i.e. exact symmetries, whose charges are integrable, the first term can be removed, and the above relation can be further reduced to
\begin{align}
\oint_{\partial\Sigma}\left(\bm{k}_{\hat{\delta}_{1}\eta}(\hat{\delta}_{2}\Phi,\Phi)-\bm{k}_{\hat{\delta}_{2}\eta}(\hat{\delta}_{1}\Phi,\Phi)\right)=\delta_2 Q_{\eta_1}-\delta_1 Q_{\eta_2}\approx 0.
\end{align}

It is easy to check that $\partial_t, \partial_\phi$ are integrable from the above relations. Let's consider their generally linear combination, 
\begin{align}
    \xi=A\partial_t+B\partial_\phi,
\end{align}
where $A$ and $B$ are some constants over spacetime but functions of the parameters of BTZ black hole, i.e. $A=A(m,j), B=B(m,j)$. Since the tangent space of $\mathcal{S}_p$, $T_{\mathcal{S}_p}$, for BTZ black hole is two dimensional and spanned by $\pdv{\Phi}{m}\delta m$ and $\pdv{\Phi}{j}\delta j$, we can just choose $\hat\delta_1\Phi=\pdv{\Phi}{m}\delta m$ and  $\hat\delta_2\Phi=\pdv{\Phi}{j}\delta j$. Then we have 
\begin{align}
    &\hat\delta_2 Q_{\xi_1}-\hat\delta_1 Q_{\xi_2}= \hat\delta_2 Q_{\pdv{A}{m}\delta m\partial_t+\pdv{B}{m}\delta m\partial_\phi}-\hat\delta_1 Q_{\pdv{A}{j}\delta j\partial_t+\pdv{B}{j}\delta j\partial_\phi}\nonumber\\ 
    =&\pdv{A}{m}\delta m\hat\delta_2 Q_{\partial_t}+\pdv{B}{m}\delta m\hat\delta_2 Q_{\partial_\phi}+\pdv{A}{j}\delta j\hat\delta_1 Q_{\partial_t}+\pdv{B}{j}\delta j\hat\delta_1 Q_{\partial_\phi}\nonumber\\ 
    =&(\pdv{A}{j}-\pdv{B}{m})\delta m\delta j,
\end{align}
in the second step, we have used the fact that the charges for the linear combination of exact symmetries are the same linear combination of the corresponding charges. We can check when $A=\frac{2\pi}{\kappa}, B=\frac{2\pi\Omega_H}{\kappa}$, the above result is vanishing, while $A=1,B=\Omega_H$, the above result is non-vanishing. The former corresponds to redefined horizon Killing vector $\zeta_H=\frac{2\pi}{\kappa}\xi_H$, while the latter just corresponds to horizon Killing vector $\xi_H$. Hence, the charge related to redefined horizon Killing vector $\zeta_H$ is integrable, while the charge related to horzion Killing vector $\xi_H$ is not integrable.

\twocolumngrid

%

\end{document}